\documentclass{acta}
\usepackage{supertabular,lscape,epsfig}
\usepackage{amssymb}
\usepackage{amsmath}
\usepackage[usenames,dvipsnames]{color}
\usepackage{url}

\SetPages{0}{0}

\SetVol{76}{2026}

\begin{document}

\begin{Titlepage}
\Title{TOI-7475 spectroscopic follow up - detection of a massive hot Jupiter\\ on an eccentric orbit\footnote{Based on the PST1 telescope spectroscopic data and photometry from TESS space telescope.} }

\Author{~J. ~P~i~e~t~r~u~c~h~a$^1$, ~J.~J.  ~T~o~k~a~r~e~k$^1$, ~M. ~W~i~t~c~z~a~k$^1$, ~W. ~D~i~m~i~t~r~o~v$^1$}
{$^1$Astronomical Observatory Institute, Faculty of Physics and Astronomy,\\ Adam Mickiewicz University, ul. S{\l}oneczna 36, 60-286 Pozna\'{n}, Poland\\
e-mail: dimitrov@amu.edu.pl\\
e-mail: joapie16@st.amu.edu.pl}

\Received{May 29, 2026}
\end{Titlepage}

\Abstract{
We present the first spectroscopic follow-up for TOI-7475, and report the detection of a planetary-mass companion. Transits with a period of 3.25 days were observed by TESS. The dips in brightness are 0.006 mag and last approximately 4.5 hours. The host star is a spectral type F5 with a mass of approximately 1.48 M$_{\odot}$.
We collected 24 high resolution echelle spectra using the PST1 telescope. Analysis of the TESS photometric data and our radial velocities was performed using the Wilson-Devinney method. We obtained a planetary mass of the companion of 12.0$\pm$1.0 M$_{\mathrm{jup}}$. The small semi-major axis and short orbital period classify the planet as a Hot Jupiter. Its orbit is elliptical with an eccentricity of 0.16. The ellipticity is evident in the spectroscopic data and could indicates the young age of the system.
}
{Exoplanet: detection, hot Jupiter}

\section{Introduction}
Among the thousands of exoplanet candidates recorded by the Transiting Exoplanet Survey Satellite (TESS) mission is TOI-7475.01. The object is listed in the NASA Exoplanet Archive  (Christiansen et al. 2025) along with other TESS targets as a planet candidate. The transit depth is 5247 ppm or approximately 0.5\%, and the orbital period is $3.252762\pm0.000053$ days. The planet orbits very close to its host star and its equilibrium temperature is estimated to be 1714 K.  The host star is also designated HD 105186 or TIC 376866659. It is known as a high proper motion star and is located approximately 61 degrees above the galactic plane. 

TOI-7475 was also observed by the GAIA\footnote{https://gea.esac.esa.int/archive/} mission, which provides data on the host  (DR 3). Its temperature is estimated at 6547 K and a surface gravity log(g) at 4.0785 (cgs). The remaining measurements are listed in Table 1.
There are five objects within a radius of 1 arcmin of TOI-7475, but all of them have significantly smaller parallaxes and proper motions, so they are background stars, dynamically unrelated to our target. The nearest object is a 14.5-magnitude star at an angular distance of $\sim$30 arcsec.

A photometric study of the object was presented by Escol'a-Rodrigo (2026). Statistical validation of the TOI-7475.01 candidate was conducted. The author confirmed the object's clean environment and estimated the probability of a false positive result (FPP) to be $\sim 0$. Based on the mass-radius relationship, the mass of the planet was estimated at $\sim 2.2$ M$_{\mathrm{jup}}$.

Since no radial velocity curve has been reported in the literature to date, we conducted a spectroscopic follow-up to determine the companion's mass and confirm whether it is a planet. The object is relatively bright (8.6 Vmag) and the expected amplitude of radial velocity variations is high; it has a short period and a deep transit. The object is suitable for our 0.5-meter telescope equipped with an echelle spectrograph.

\MakeTable{cccccc}{12.5cm}{GAIA DR3 measurements for TOI-7475 host star.}
{
\hline
\vspace{1mm}
       Parallax (mas)       & Distance (pc)*        & BP (mag)     & BP-RP (mag)    & RV ($km~s^{-1}$)    &  Fe/H   \\
\hline\noalign{\smallskip}
$8.358\pm0.034$     &  $119.09$  & $ 8.699$    & $ 0.606$     & $38.08\pm0.31$     &    $-0.332$  \\

\noalign{\smallskip}\hline
* photo geometric distance
}

\section{TESS photometry}
Our model of TOI-7475 is based on high-precision photometry from the TESS space telescope. The TESS lightcurves were obtained using the Science Processing Operations Center (SPOC; Jenkins et al. 2016)\footnote{https://archive.stsci.edu/hlsp/tess-spoc} pipeline and were downloaded from the MAST archive. The data came from TESS sector 91 and were collected between April 9 and May 7, 2025. The photometric data were extracted using the TOPCAT software. For further analysis, we selected the columns containing the observation time, the Simple Aperture Photometry flux and its corresponding measurement error. The timestamps, originally in the Barycentric TESS Julian Date (BTJD) format, were than corrected to the Barycentric Julian Date (BJD) standard by adding 2 457 000 days. Finally, to obtain normalized light curves, the flux values and their associated errors were divided by the median flux of the dataset. Slow trends were removed using a rolling-median filter applied separately to continuous observing segments, with the window length set equal to the orbital period of the planet. The light curve was then divided by this trend with masked-off transits.
Since there were about 90,000 observation points, the data were aggregated into normal points by averaging 10 or 100 points together for modeling purposes.
Equation 1 shows the ephemeris we used in our subsequent analysis.

\begin{eqnarray}
{\rm BJD_{\rm min}} & = & 2460775.586536 + 3.252762 \times E,\\
&& \quad \pm 0.000242 \qquad \pm 0.000053 \nonumber
\label{Eq1}
\end{eqnarray}

\section{PST1 spectroscopy}
Our spectroscopic data was collected with the PST1 telescope. The spectra were obtained from 17 March to 3 May 2026. The telescope has a relatively small diameter of 0.5 m, but thanks to a good fit with the optical 
fiber and spectrograph it has low light loss and allows observing objects up to 11 mag. The telescope was first described by Baranowski et al. (2009). 
The \'{e}chelle spectrograph has a resolution R $\sim$ 35000. Python/IRAF scripts were used to reduce the spectroscopic data. Velocity measurements were performed using the cross-correlation method; we used the FXCOR IRAF task.
The measured velocities show a peak-to-peak variation of 2.5$km~s^{-1}$. After phasing with the transit period, we obtained a non-sinusoidal radial velocity curve, indicating an elliptical orbit. The 1-sigma standard deviation of the data is about 70 $m~s^{-1}$.

\MakeTable{ccc}{12.5cm}{Radial velocity measurements for TOI-7475 host star obtained with cross-correlation\\ method (FXCOR) and converted to the barycentric reference frame.}
{\hline
   ~~~~~~~~~~BJD$_{\mathrm{TDB}}$ ~~~~~~~~ &  RV  & ~~~~~~~~error~~~~~~~~ \\
                                  &  $km~s^{-1}$   &  $km~s^{-1}$  \\
\hline\noalign{\smallskip}

 2461117.42633  &   37.624  &  0.209\\
 2461117.44883  &   37.750  &  0.270\\
 2461121.39393  &   37.057  &  0.151\\
 2461121.41642  &   36.910  &  0.173\\
 2461121.54111  &   37.112  &  0.174\\
 2461121.56361  &   37.129  &  0.174\\
 2461127.41206  &   37.512  &  0.144\\
 2461127.43455  &   37.484  &  0.165\\
 2461132.37442  &   38.793  &  0.208\\
 2461132.39721  &   38.773  &  0.238\\
 2461142.34701  &   39.572  &  0.187\\
 2461142.44634  &   39.734  &  0.180\\
 2461142.46883  &   39.263  &  0.213\\
 2461148.44745  &   38.487  &  0.167\\
 2461148.46994  &   38.584  &  0.174\\
 2461152.42764  &   39.572  &  0.211\\
 2461152.45013  &   39.598  &  0.160\\
 2461162.38544  &   39.128  &  0.248\\
 2461162.40793  &   39.241  &  0.193\\
 2461163.33628  &   37.350  &  0.184\\
 2461163.40358  &   37.184  &  0.148\\
 2461163.42607  &   37.230  &  0.157\\
 2461164.34072  &   37.652  &  0.339\\
 2461164.36819  &   37.656  &  0.253\\

\noalign{\smallskip}\hline}
\clearpage

\section{Modeling of the system}
For modeling, we used the Wilson-Devinney method (1971; Wilson 1990, 1994) and the PHOEBE\footnote{https://phoebe-project.org/} code (Pr{\v s}a and Zwitter 2005). We analyzed the light curve from the TESS satellite and the radial velocity curve obtained using PST1. We obtained data on the host star from GAIA. As a first approximation, we assumed that it is on the main sequence based on log(g) (at lower limit for the main sequence) and the fact that the orbit is elliptical, which indicates a young age for the system. Using the tables from Pecaut \& Mamajek (2013)\footnote{https://www.pas.rochester.edu/\textasciitilde{}emamajek/EEM\_dwarf\_UBVIJHK\_colors\_Teff.txt}
and the temperature from GAIA DR3 of 6547 K, we obtained a mass of 1.33 M$_{\odot}$. We assumed synchronous rotation and used Van Hamme's (1993) limb darkening coefficients.

Preliminary fits confirmed that the radius of the parent star is larger than that of a main-sequence star. This means that we have underestimated the star's mass. 
We found that for a slightly higher mass, we obtain values for log(g) and temperature that are consistent with observations.
For a mass of 1.48 M$_{\odot}$, we obtained a log(g) of 4.078 and a radius of 1.84 R$_{\odot}$ at a temperature of 6550 K.
A standard star with similar parameters is Procyon ($\alpha$ CMi); its spectral type is F5IV-V.
In the second attempt, we used that mass value for the parent star in our modeling.
This mass yields a much more consistent result. The planet's mass is then 12.0$\pm$1.0 M$_{\mathrm{jup}} $.
The radius of the planet is $1.23\pm0.04~$R$_{\mathrm{jup}}$, and its density is 6.45 times higher than that of Jupiter.

The radial velocity curve clearly indicates the ellipticity of the orbit. However, the light curve shows a shift in phase of about 0.02. This cannot be explained by an incorrect period, because a period that fits both curves well causes the transit to be blurred. Since the cause is not the orbital period, the possibility of apsidal motion remains. Less than a year, or about 100 orbital cycles, elapsed between the photometric and spectroscopic data. Simultaneous fitting of LC/RV did not yield a satisfactory solution -- we were unable to find an orbital orientation that would be consistent with both curves. We therefore conclude that the discrepancy is due to apsidal motion.
In the case of the radial velocity curve, the solution converges quickly and is unambiguous. In the case of the light curve, only the main transit is visible, so the argument of periastron and the phase shift cannot be unambiguously determined.
The results of the final fitting are given in Table 3. Our systemic velocity is in good agreement with the value obtained by GAIA (38.08 $km~s^{-1}$, Tab. 1). 
The flat-bottomed transit and the fitted model are shown in Figure 1. The flux is normalized to 1. The model reproduces the ingress and egress slopes, as well as the brightness changes during the total phase.
The radial velocity plot is also centered on phase zero (Figure 2). The model predicts a Rossiter-McLaughlin effect with semi-amplitude of about 100 $m~s^{-1}$ for synchronous rotation.

\begin{figure}[h!]
\begin{center}
\vspace{1.5cm}
\includegraphics[width=0.9\textwidth]{LC.eps}
\vspace{1mm}
\FigCap{The final model compared with photometric data zoomed in on the transit. The synthetic curve is shown as a solid orange line. The original data are represented by gray dots, and the smoothed data (averaged every 10 points) are shown as blue circles.}
\end{center}
\end{figure}

\begin{figure}[h!]
\begin{center}
\vspace{1.5cm}
\includegraphics[width=0.9\textwidth]{RV.eps}
\vspace{1mm}
\FigCap{Radial velocity plot for TOI-7475. The blue circles represent our velocity measurements along with their errors. The solid line represents our final model; the Rossiter effect is visible near the zero-phase point during the transit. The ellipticity of the orbit is clearly visible.
}
\end{center}
\end{figure}

\MakeTable{lcc}{12.5cm}{Parameters obtained for TOI-7475 system. The planet's equilibrium temperature is given both with and without thermal redistribution, assuming an albedo of 0.1.}
{\hline
                                 &    Star    &    Planet       \\
\hline\noalign{\smallskip}
 $i$                               & \multicolumn{2}{c}{$87.\!\!^{\circ}18 \pm 0.03$}    \\ 
 $q$                               & \multicolumn{2}{c}{$0.00772 \pm 0.00001$}               \\ 
 $a~($R$_{\odot})$               & \multicolumn{2}{c}{$10.561  \pm 0.293$}               \\ 
 $a~(\mathrm{AU})$                          & \multicolumn{2}{c}{$0.04914  \pm 0.00136$}               \\
 $V_{\gamma}$~(km~s$^{-1}$)      & \multicolumn{2}{c}{$38.16  \pm 0.03$   }               \\ 
&&\\
 $e$                                & \multicolumn{2}{c}{$0.155 \pm 0.033$}               \\ 
 $\omega$                         &\multicolumn{2}{c}{$6.240 \pm 0.209$}               \\ 
$ph_{shift}$                      & \multicolumn{2}{c}{$0.0709 \pm 0.0058$}               \\ 
&&\\
 $T$~(K)                        & $6550\pm150$     &  1330 / 1590     \\
 $\Omega$                      & $5.7377 \pm 0.0171$ &  $2.3228  \pm 0.0026$  \\
&&\\
 $M ~($M$_{\odot})$           & $1.482 \pm 0.130$  & $0.0114  \pm 0.0010 $  \\
 $M ~($M$_{\mathrm{jup}})$             &                     & $ 12.0 \pm 1.0 $  \\
&&\\
 $R ~($R$_{\odot})$           & $1.847  \pm 0.058$  & $0.127  \pm 0.004 $  \\
 $R ~($R$_{\mathrm{jup}})$             &                     & $ 1.23 \pm 0.04 $  \\
&&\\
 $log(g)$ (cgs)               & $4.076    \pm 0.015$ & $4.291  \pm 0.016$   \\
\noalign{\smallskip}\hline

}

\clearpage


\newpage
\section{Conclusions and discussion}
Our follow-up observations indicate that TOI-7475 hosts a massive hot Jupiter in an eccentric orbit. TOI-7475b is a gas giant strongly heated by its host star. The amplitude of the radial velocity curve is high, and the shape is clearly non-sinusoidal. 
The evolutionary status of the host star must be clarified; the star's radius is larger than we would expect on the main sequence. It is a pre-main-sequence star or is approaching the subgiant phase. The elliptical orbit of the planet and the strong tides suggest a young age for the system.
To determine the parameters of the planet, we need the most accurate values of
the parameters of the parent star. We are continuing our spectroscopic observations, on the one hand to refine the orbital solution and, on the other hand, to perform a spectral analysis and more accurately determine the host star's surface temperature and surface gravity log(g). Future observations will also allow us to confirm or rule out apsidal motion.
Variability outside the transit must be analyzed; its flux amplitude is on the order of 0.001, it is irregular, and it is likely related to the stellar variability. The model predicts ellipsoidal brightness variations with an amplitude of 0.00015 well below the stellar variability. The secondary transit is not visible, suggesting a low planetary albedo.
TOI-7475b is one of the hot Jupiters with the highest amplitude of radial velocity variations; in this group, WASP 18b (Southworth et al. 2009) holds the record with a semi-amplitude of 1.8 $km~s^{-1}$.
The object we are studying also has a deep transit of 0.5\% and is relatively bright (8.6 mag), making it an ideal candidate for the future precise determination of the planet's parameters.

\Acknow{
This work has made use of data from the Transiting Exoplanet Survey Satellite (TESS) and Pozna\'{n} Spectroscopic Telescope 1 (PST1).
The TESS data were obtained from the Mikulski Archive for Space Telescopes (MAST) using Science Processing Operations Center (SPOC) pipeline.

This work has made use of data from the European Space Agency (ESA) mission Gaia (https://www.cosmos.esa.int/gaia), processed by the Gaia Data Processing and Analysis Consortium (DPAC, https://www.cosmos.esa.int/web/gaia/dpac/\\consortium). Funding for the DPAC has been provided by national institutions, in particular the institutions participating in the Gaia Multilateral Agreement.

The authors are grateful to Tomasz Kwiatkowski, Przemys\l{}aw Bartczak, Aleksander Schwarzenberg-Czerny and our engineer Roman Baranowski, 
founders of the Pozna\'{n} Spectroscopic Telescope project.
We thank Wiktoria Stefanowska and Pawe\l{} Bia\l{}ob\l{}ocki for their contribution to the observations and Olgierd Jastrz\k{e}bski and Mi\l{}osz Strzelczyk for their help with the photometric data.
}

\end{document}